\documentclass[%
 reprint,
 amsmath,amssymb,
prl,
]{revtex4-2}
\usepackage{amsmath}
\usepackage{graphicx}
\usepackage{dcolumn}
\usepackage{bm}
\usepackage {verbatim}
\usepackage{color}

\begin{document}

\preprint{APS/123-QED}

\title{Isospectral reduction of the SSH3 lattice and its bulk-edge correspondence}

\author{Qian-Hao Guo, Yang Zhang, Xiao-Huan Wan,}
\affiliation{%
	School of Science, Shenzhen campus of Sun Yat-sen University, 518107, Shenzhen, China}%
\author{Li-Yang Zheng}%
 \email{zhengly27@mail.sysu.edu.cn}
\affiliation{%
School of Science, Shenzhen campus of Sun Yat-sen University, 518107, Shenzhen, China}%

\date{\today}

\begin{abstract}
Here, we propose an isospectral reduction (IR) approach for the mapping of a trimer Su-Schrieffer-Heeger (SSH3) lattice into a simplified two-site model, whose coupling dynamics ingeniously results in a precise bulk-edge correspondence of the original lattice. The isospectrally-reduced model has inter-cell couplings with dynamic response to the eigenstate energy, allowing for the control of topological phase transition by energy. We relate the bulk property of the reduced model to the band topology of the SSH3 lattice, allowing for there distinct topological phases with different number of topological edge state pair. An acoustic SSH3 chain is fabricated for experimental demonstration. The said topological edge state pairs are measured. Our study takes a pivotal step toward the exploration of topology in multiple wave systems, opening up possibilities for advanced control of topological waves.
\end{abstract}

\maketitle

The study of topological properties has attracted worldwide attention in recent years \cite{PhysRevLett.114.114301,ye2022topological,ozawa2019topological,liao2024visualizing,zhang2023second} . 
Robust transport of topological edge states is one of the most appealing properties as these edge states exhibit immunity to certain defect/disorder \cite{PhysRevApplied.12.034014,zhang2021superior,jin2018robustness,wang2021structural}. The bulk-edge correspondence that links the number of edge states to a topological invariant is being the central issue in topological systems \cite{graf2013bulk,kawarabayashi2021bulk,mong2011edge,asboth2016short}. 
For instance, the quantum Hall insulators, known as Chern insulators supporting chiral edge states, are characterized by a Chern number~\cite{thouless1982quantized,haldane1988model,xue2022topological}. 
Likewise, the spin Hall insulators or $Z_2$ topological insulators exhibiting a pair of helical edge states are protected by a quantized $Z_2$ topological invariant \cite{kane2005quantum,susstrunk2015observation,he2016acoustic,song2015quantization}. 
In general, the bulk-edge correspondence and the robustness of topological states require specific symmetries (e.g., time-reversal, chiral or spatial symmetry) such that the topological invariant is quantized. However, 
recent developments reveal exotic topological phases of matter that go beyond the conventional scenario, leading to topological states in non-Hermitian \cite{gong2018topological,okuma2023non,lieu2018topological}, nonlinear \cite{zangeneh2019nonlinear,tuloup2020nonlinearity} and non-Abelian systems \cite{jiang2021experimental,guo2021experimental}. Even in a simple periodic structure of the SSH3 lattice, a model lacking apparent lattice symmetry due to odd number of sublattices, is reported to have topological edge states~\cite{PhysRevA.99.013833,Anastasiadis_2022,eiles2024topological,Xie_2019}.  
More interestingly, the signature of topological phase transition indicated by closing and reopening a bandgap is unnecessarily shown in the SSH3 model when its band topology is altered. This brings difficulties to define a bulk invariant capable of counting the edge state number according to the topological classification with symmetry \cite{chiu2016classification,wang2020construction}. 

In viewing of establishing the bulk-edge correspondence in the SSH3 model, a sequence of attempts have been made based on Green’s function \cite{PhysRevA.99.013833}, sublattice Zak’s phase \cite{Anastasiadis_2022,sougleridis2024existence} or Chern number in a synthetic dimension \cite{eiles2024topological}. 
Instead, we address the essential issue for generalization of mapping between the SSH3 model to an equivalent but simplified one, through which the bulk invariant of the SSH3 model can be easily characterized and its connection to the number of edge states can be found. 
The study of isospectral reduction (IR) rising from graph theory might bring the solution. Mathematically, IR is a method of reducing a graph (described by an adjacency matrix) to a graph with fewer vertices whose adjacency matrix has the same eigenvalues and eigenvectors~\cite{kempton2020characterizing,smith2019hidden}. This method has been used to create stability preserving transformations of networks~\cite{bunimovich2011isospectral,rontgen2023hidden}, and study the survival probabilities in open dynamical systems~\cite{bunimovich2014improved}. Recent studies of topological states show that IR of the system's Hamiltonian can be used to unearth the hidden symmetry, giving rise to exotic band degeneracy \cite{hou2018hidden,li2024group,rontgen2021latent} and robust topological edge states \cite{rontgen2024topological,zheng2023robust}.  

In this work, we show how the IR approach allows us to map the SSH3 lattice into an isospectrally-reduced model (IRM) without altering its eigenspectrum. Based on the IRM, whose coupling dynamics is described by a nonlinear two-site Hamiltonian, we show how the energy band of the SSH3 model is related to the bulk topology of the IRM, and how the bulk-edge correspondence of the SSH3 model is established from the IRM Hamiltonian. Our work provides a simple strategy for the study of topological property in systems beyond the tenfold way classification of topological insulators and superconductors\cite{sa2023symmetry,chen2019free,PhysRevB.78.195125}, bringing possibilities for the exploration of topological wave effects in multiple wave systems.

\begin{figure*}[t]
	\includegraphics[width=1\textwidth]{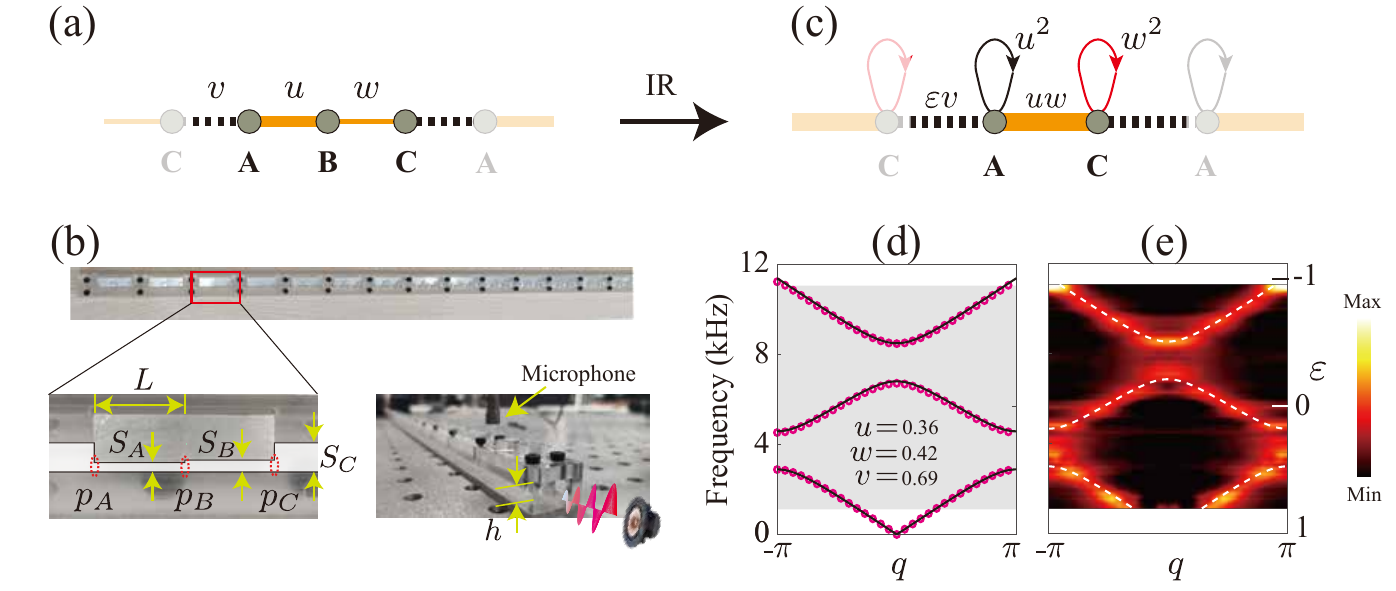}
	\caption{\label{fig:11} 
(a) Schematic presentation of the SSH3 model.
(b) A sample of the acoustic SSH3 chain and experimental setup. The zoomed view of the a unit cell is presented in the bottom left panel. The bottom right panel displays a picture of the measurement system.
(c) Graphical representations of the IRM of the SSH3 model. Arrowed lines stand for the on-site couplings.
(d) Dispersion curves of the SSH3 chain. Solid lines represent theoretical calculation and red circles correspond to simulation result. 
(e) Measured band structure. Color level indicates the mode intercity. Dashed lines are the theory prediction.}
\end{figure*}

The SSH3 model is shown in Fig.~\ref{fig:11}(a) where the unit cell contains three sites $A$, $B$ and $C$ interacting through the intra-cell couplings $u$, $w$ (orange) and the inter-cell coupling $v$ (black dashed). The Hamiltonian reads
\begin{eqnarray}\label{eq:1}
H(q)=
\begin{pmatrix}
 	0 & u & v{{e}^{-iq}}  \\
 	u & 0 & w  \\
 	v{{e}^{iq}} & w & 0  \\
\end{pmatrix}.
\end{eqnarray}
$q\in [-\pi, \pi]$ is the normalized wave vector. The SSH3 model can be realized in acoustics using connected tubes with alternating cross sections. A sample of $31$ cells made of plexiglass is fabricated in Fig.~\ref{fig:11}(b), where a unit cell is highlighted in the zoomed view. 
As seen, three air channels (white area) of the same length $L$ and height $h$ but different widths $S_A,S_B$, and $S_C$ are connected to form an acoustic SSH3 chain. Under the low-frequency approximation, i.e., $S_A,S_B, S_C,h \ll L$, sound propagation in the chain can be described by an eigenvalue problem, $H(q) X= \varepsilon X$ with $X=[\tilde{p}_A; \tilde{p}_B; \tilde{p}_C]=[\sqrt{\alpha }p_A;\sqrt{\beta }p_B;\sqrt{\gamma }p_C]$ composed of pressures at the three junctions. $\varepsilon$ represents the eigenenergy, and $H(q)$ has the exact form of Eq.~\ref{eq:1} under the bridging relations~\cite{zheng2020acoustic,coutant2021acoustic},
\begin{equation}\label{eq:2}
u\rightarrow \dfrac{S_A}{\sqrt{\alpha \beta }},\ w \rightarrow \dfrac{S_B}{\sqrt{\beta \gamma}},\ v \rightarrow \dfrac{S_C}{\sqrt{\alpha \gamma }},\ \varepsilon\rightarrow \cos \dfrac{2\pi f L}{c}.
\end{equation}
Above, $\alpha={{S}_{A}}+{{S}_{C}},\beta={{S}_{A}}+{{S}_{B}},\gamma={{S}_{B}}+{{S}_{C}}$. $f$ and $c$ represent frequency and speed of sound, respectively. 
By setting $S_A = 2 \text{ mm}$, $S_B = 2.4 \text{ mm}$, $S_C = 5 \text{ mm}$, $h = 5 \text{ mm}$, $L = 15 \text{ mm}$, we have $u=0.36, w=0.42, v=0.69$. The dispersion curves of $H(q)$ are shown in Fig.~\ref{fig:11}(d) by black lines, showcasing good agreement with the simulation (red circles). In experiment, a perforation (radium $4$ mm) is designed at each node and is left opened only when the pressure at that node is being recorded by a microphone, Fig.~\ref{fig:11}(b). A loudspeaker connected to a function generator is placed closely at one end to launch sound waves in the chain. 
By spatially and spectrally collecting the measured data, the measured band structure is obtained after a Fourier transform. As depicted in Fig.~\ref{fig:11}(e), good agreements between prediction (dashed lines) and experiment (the color level indicates the mode intensity) are observed.
 
We now introduce the IRM of the SSH3 chain. 
Consider nodes $A$ and $C$ as the `fringe sites' having both intra-/inter-cell couplings, and node $B$ as the `inner site' with pure intra-cell couplings only, this treatment allows us to isospectrally reduce the three-site Hamiltonian $H(q)$ into the fringe subspace $S =\{ {\tilde p_A},{\tilde p_C}\}$ in which the inner degree of freedom $\bar S =\{ \tilde p_B\}$ is completely integrated out. Thus, sound wave dynamics in $S$ can be described by a nonlinear eigenvalue problem 
\begin{eqnarray}  	\label{eq:3}
\mathcal{H}(\varepsilon, q)
\left( \begin{array}{l}
	\tilde p_A\\
	\tilde p_C
\end{array} \right)
=
{\varepsilon ^2}\left( \begin{array}{l}
	\tilde p_A\\
\tilde p_C
\end{array} \right),\ 
\mathcal{H}=\varepsilon\mathcal{R}_{S}=
\left( {\begin{array}{*{20}{c}}
		{{u^2}}&{\rho}\\
		{\rho^*}&{{w^2}}
\end{array}} \right),
\end{eqnarray} 
where $\mathcal{R}_S=H_{SS}-H_{S\bar S}(H_{\bar S \bar S}-\varepsilon I)^{-1}H_{\bar S S}$ is the IR matrix of $H(q)$ over $S$, and $\rho= uw +\varepsilon v e^{-iq}$.   
Interestingly, when regarding $\mathcal{H}$ as an effective Hamiltonian, Eq.~\eqref{eq:3} describes the couplings of a two-site model shown in Fig.~\ref{fig:11}(c), demonstrating that it is the IRM of the SSH3 lattice since they have the same eigenspectrum. 
Due to the inner degree of freedom in the SSH3 chain, the on-site coupling $u^2$ ($w^2$) is allowed via the interaction path $A (C)\rightarrow B \rightarrow A (C)$. And the path $A (C)\rightarrow B \rightarrow C (A)$ between sites $A, C$ leads to the intra-cell coupling $uw$ in the IRM.  
Note that, the strength of the inter-cell coupling $\varepsilon v$ of the IRM depends on the eigenenergy $\varepsilon$, reflecting the tunability of dynamical response to different eigenstates of the SSH3 lattice. 
In other words, the eigenstates in different energy bands of the SSH3 lattice can have distinct coupling response in the IRM, resulting in the possibilities of topological phase transition in the IRM when the ratio of the intra-cell coupling to the inter-cell one $\left|\frac{uw}{v\varepsilon}\right|$ is shifted from larger to smaller than $1$.  

\begin{figure}[t]
	\includegraphics[width=0.48\textwidth]{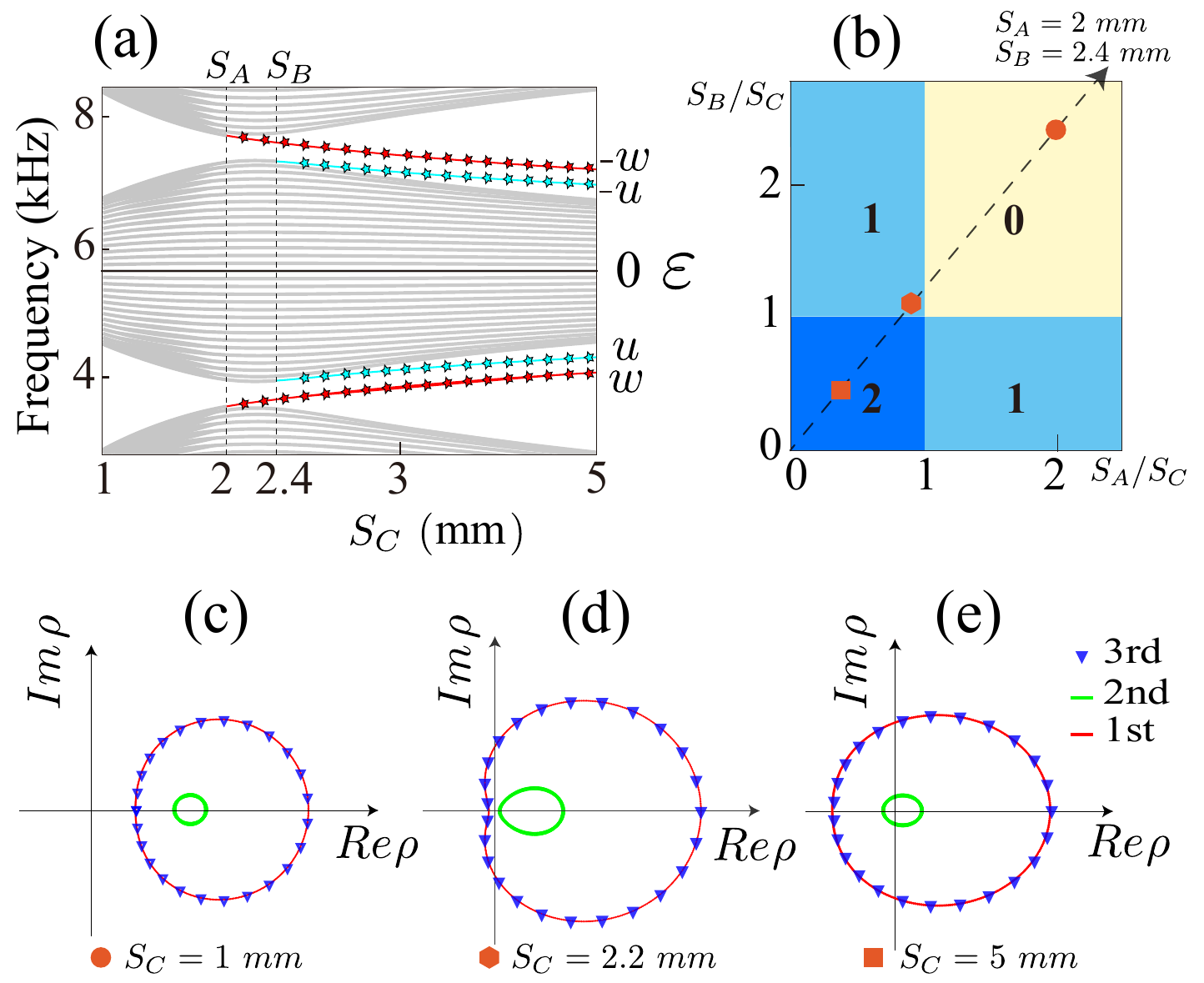}
	\caption{\label{fig:22} 
		(a) Eigenspectra of the chain with 31 cells in dependence of $S_C$ by fixing $S_A=2$ mm and $S_B=2.4$ mm.
		(b) Phase diagram for different $S_C$. The yellow zone has trivial phase. The chain supports one pair of edge states in the light blue zones, while it has two pairs for the phase in the deep blue zones. 
		(c)-(e) Windings of $\rho$ for the three cases marked in (b).}
\end{figure}

When the IRM exhibits nontrivial topology, the corresponding SSH3 chain supports topological edge state pairs at the chain terminations. In Fig.~\ref{fig:22}(a), we present the change of eigenspectra of the finite SSH3 chain when $S_C$ is increased while $S_A=2$ mm, $S_B=2.4$ mm are fixed. Gray lines correspond to bulk states. As seen, when $S_C<S_A$, the chain has zero edge state, while there exist a pair of edge states (red lines) in the bandgaps (one edge state in each bandgap) starting from $S_C>S_A$. When $S_C>S_B$, another pair of edge states appear (cyan lines). In this process, there is no sign of bandgap closing and reopening. In fact, the topological phase transition can be characterized by the term $\rho$ in the IRM which does not require the closing of the banggap.

The IRM Hamiltonian $\mathcal{H}$ in Eq.~\ref{eq:3} indicates that when $\rho=0$, the system exhibits an on-site inversion symmetry $\Sigma=[1,0;0,-1]$ satisfying $[\mathcal{H}, \Sigma]=0$. This means that the eigenstates $\phi=[1;0]$ and $\psi=[0;1]$ of $\Sigma$ are also the eigenstates of $\mathcal{H}$. Their corresponding eigenvalues are $\varepsilon_\phi^2=u^2$ and $\varepsilon_\psi^2=w^2$. In other words, if the eigenstate $\phi$ ($\psi$) exists, it must appear in pair at $\varepsilon_{\phi}^\pm=\pm u$ ($\varepsilon_{\psi}^\pm=\pm w$). 
Consider a complex edge wavevector $q=\kappa-i\zeta$, the condition $\rho= 0$ leads to a pair of edge wavevectors 
\begin{eqnarray} \label{eq:4}
\begin{array}{l}
\kappa_j^-=2n\pi,\ \ \ \ \ \ \ {\zeta}_j^- =-\ln \left |\frac{uw}{v\varepsilon_j^-}\right|; \\
\kappa_j^+=(2n+1)\pi,\ \ {\zeta}_j^+ =-\ln \left |\frac{uw}{v\varepsilon_j^+}\right|.  	
\end{array}
\end{eqnarray}
$n$ is an integer, and $j=\phi,\psi$. The evanescent nature of edge state requires $\zeta>0$, namely $\left|\frac{uw}{v\varepsilon}\right|<1 $. 
Thus, we conclude that there always exist a pair of eigenstates localized at the chain terminations but decaying along the chain that is protected by the on-site inversion symmetry of the IRM as long as $\rho=0$ (i.e., $\left|\frac{uw}{v\varepsilon}\right|<1 $). 
Their wavefunctions can be expressed as
\begin{eqnarray} \label{eq:5}
\begin{array}{l}
\phi_l^{\pm}= \pm
\left( \begin{array}{l}
1\\
0
\end{array} \right) e^{-(l-1)\zeta^{\pm}_\phi},\ 
\psi_l^{\pm}= \pm
\left( \begin{array}{l}
0\\
1
\end{array} \right) e^{-(l-1)\zeta^{\pm}_\psi}. 
\end{array}
\end{eqnarray}
$l=1,..,N$ for a chain of $N$ cells. For the edge state pair of $\phi$ ($\psi$) type, they appear at $\varepsilon_\phi^\pm=\pm u$ ($\varepsilon_\psi^\pm=\pm w$), giving rise to the phase condition $v>w$ ($v>u$) after substituting into $\left|\frac{uw}{v\varepsilon}\right|<1 $. 
The topological phase diagram is shown in Fig.~\ref{fig:22}(b). 
In the phase of $v<u, w$ ($S_C<S_A,S_B $), the bulk topology (yellow zone) of the IRM is trivial with zero edge states. In the light blue zones where $w<v<u$ ($S_B<S_C<S_A$) or $u<v<w$ ($S_A<S_C<S_B$), either the $\phi$ pair or the $\psi$ pair appear in the bandgaps. When $v>u, w$ ($S_C>S_A,S_B $), the IRM is transformed into a new topological phase (deep blue zone), supporting both the $\phi$ and $\psi$ pairs. 

\begin{figure*}[t]
	\includegraphics[width=0.9\textwidth]{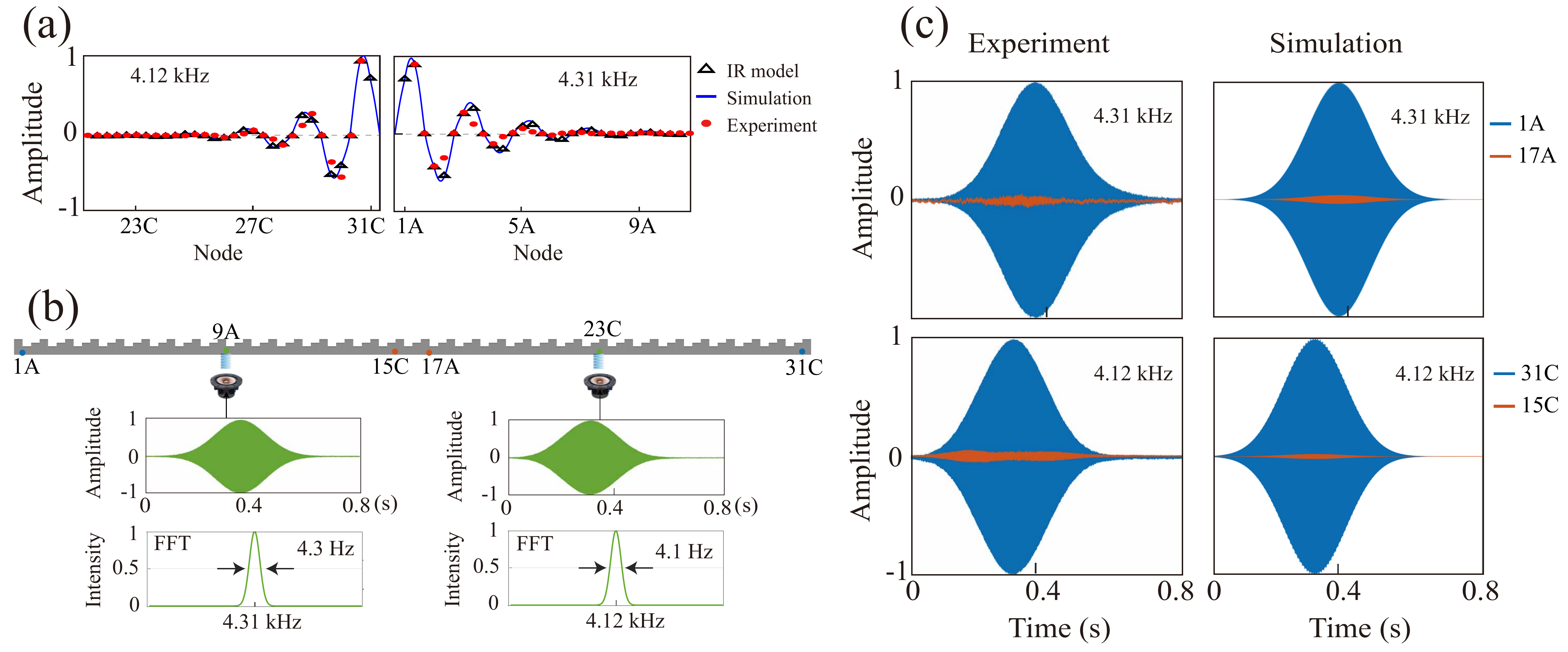}
	\caption{\label{fig:33} 
		(a) Eigen-profiles of two edge states in the first bandgap.
		(b) Schematic of experimental setup for localization demonstration of edge states. A Gaussian packet with central frequency around 4.31 kHz (4.12 kHz) is used as input imposing onto the 9A (23C) node. The pressures at the 1A and 17A (31C and 15C) are measured in (c).
	}
\end{figure*}

The bulk topology of the IRM enforces a bulk-edge correspondence for the SSH3 chain that can be characterized by the winding number of $\rho$. Consider the three cases marked in the dash line with $S_A=2$ mm and $S_B=2.4$ mm in Fig.~\ref{fig:22}(b), 
the windings of $\rho$ for different $S_C$ are shown in Figs.~\ref{fig:22}(c)-(e), where the red (green) curves correspond to the trajectories of the first (second) band for $q\in [-\pi, \pi]$, and the ones of the third band marked by blue triangles overlap with the first band. As seen, the curves of $\rho$ may or may not wind around the origin, leading to completely distinct topology consequence. 
(1) In the trivial phase, the windings of $\rho$ when $S_C=1$ mm are zero for all the bands. The total winding number $\nu=0$ counts zero edge states. (2) For the case of $S_C=2.2$ mm, the winding of $\rho$ is 1 for the first and the third bands but 0 for the second band, leading to $\nu=2$ that counts for the $\psi$ pair of edge states at $\varepsilon_\psi^\pm=\pm w$. (3) When $S_C=5$ mm, we have $\nu=4$ because the second band in this case winds twice around the origin, and the first and the third bands remain with 1. Thus, the $\phi$ pair of edge states at $\varepsilon_\phi^\pm=\pm u$ also appear. 

The existence of the $\phi$ and $\psi$ edge states can be observed in the acoustics SSH3 chain. A source sending sinusoidal wave at $\sim 4.12$ kHz ($\varepsilon \simeq w$) [or at $\sim 4.31$ kHz ($\varepsilon \simeq u$)] is located at the right (or at the left) end of the chain. Sound pressures starting from the $31$th (or the 1st) $B$ node (noted as 31B or 1B) site are measured. The measured edge wave profiles (red dots) are shown in Fig.~\ref{fig:33}(a). 
Using the relation $\varepsilon \tilde p_B=u\tilde p_A+w\tilde p_C$, the edge wavefunctions of the SSH3 chain in the first bandgap ($\varepsilon_j^+$) can be analytically obtained, 
\begin{eqnarray} \label{eq:6}
\Phi_l(\varepsilon_\phi^+)
=
\left( \begin{array}{l}
1\\
1\\
0
\end{array} \right) e^{-(l-1)\zeta^{+}_\phi},\  
\Psi_l(\varepsilon_\psi^+)
=
\left( \begin{array}{l}
0\\
1\\
1
\end{array} \right) e^{-(l-1)\zeta^{+}_\psi}. 
\end{eqnarray}
After the transform $[{p}_A; {p}_B; {p}_C]=[\frac{\tilde p_A}{\sqrt{\alpha }}; \frac{\tilde p_B}{\sqrt{\beta }}; \frac{\tilde p_C}{\sqrt{\gamma }}]$,  
the predicted edge wave profiles marked by triangles are also displayed in Fig.~\ref{fig:33}(a) ($\Psi_l$/$\Phi_l$ in the left/right panels). The simulation results are marked by the blue lines. 
It shows that the measured results have good agreements with the IR prediction and simulation. And the fact that $p_A=0$ ($p_C=0$) for all the $A$ ($C$) sites confirms that this edge state is the $\psi$ ($\phi$) type as expected.
To further verify the localized property of edge states, a sinusoidal signal $\sim 4.31$ kHz ($\sim 4.12$ kHz) modulated by a Gaussian pulse is used to launch sound from the $9$A ($23$C) node. The time signal and the corresponding Fourier transform are shown in (b).
We record the pressures of the $1$A and $17$A (the $15$C and $31$C) nodes in a duration of $0.8$ s. The measured signals are presented in Fig.~\ref{fig:33}(c), where the sharp difference in amplitude (normalized to the maximum) between nodes $1$A ($31$C) and $17$A ($15$C) are observed. This indicates that sound propagation in the chain is prohibited, while the corresponding edge state are excited, leading to the localization of sound on the edges of the chain. Same processes in simulation are also depicted in Fig.~\ref{fig:33}(c), exhibiting good agreements with the measurements.    

\begin{figure}
	\includegraphics[width=0.48\textwidth]{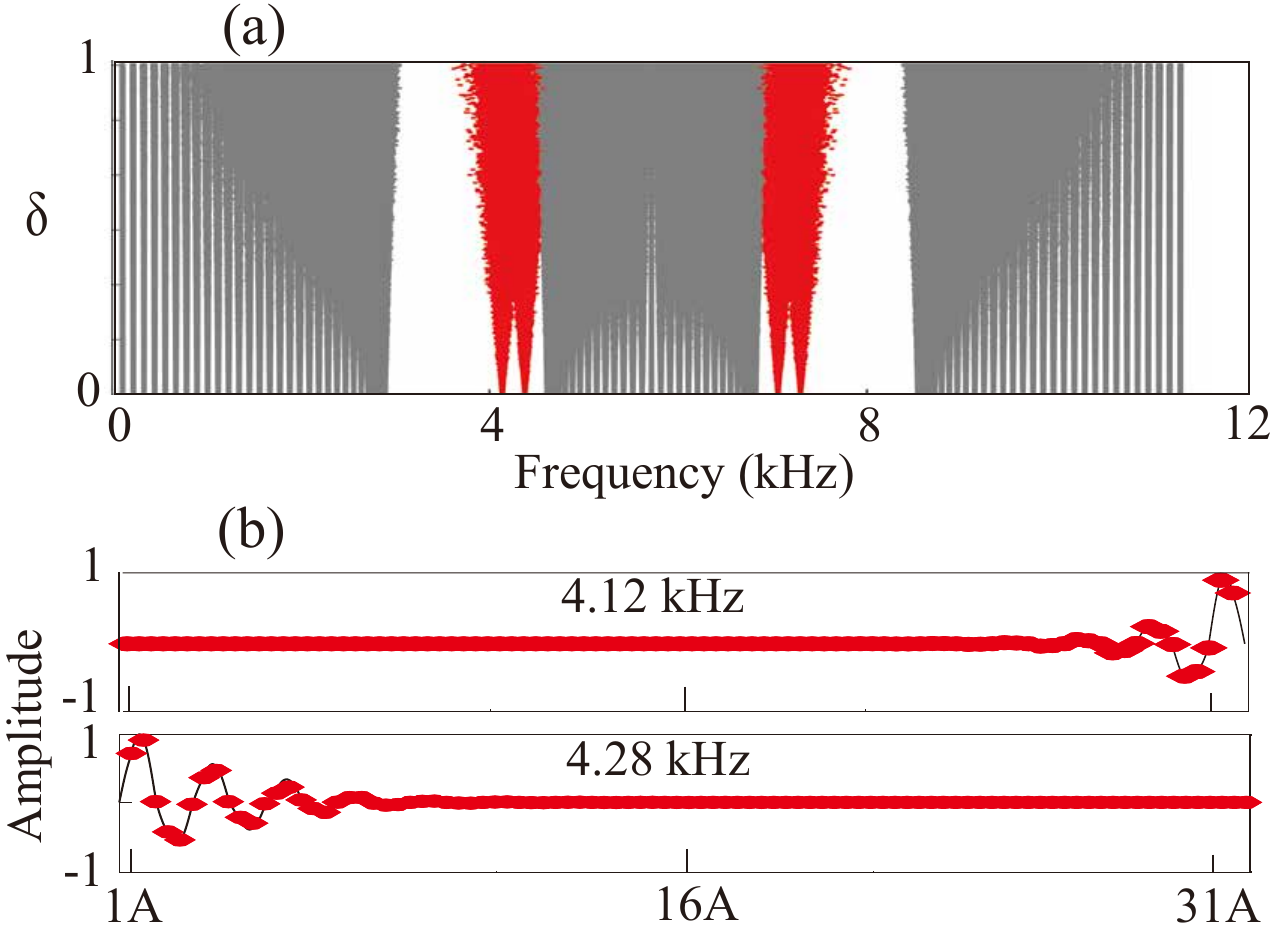}
	\caption{\label{fig:44} (a) Robustness of topological edge states against disorder. For each perturbation $\delta$, the channel widths are randomly ranged in ${S_i} = [{S_N} - \delta ,{S_i} + \delta]$ with $i=A,B,C$. Gray/red dots represent the bulk/edge states. (b) Two eigen-profiles of the edge states in the first bandgap when $\delta=1$ mm. Red dots/black lines are the theory/simulation results.}
\end{figure}
The topological edge states exhibit robustness against disorders. A perturbation $\delta \in [0,1]$ mm is now imposed on the channel widths of the sample in Fig.~\ref{fig:11}(c). For different channel along the chain, its original width $S_i$ with $i=A,B,C$ is randomly perturbed to a value ranged in $(S_i-\delta, S_i+\delta)$. However, this disorder caused by $\delta$ does not changed the fact that $S_C\in [4,6]$ mm remains larger than $S_A\in [1,3]$ mm and $S_B\in [1.4,3.4]$ mm, ensuring the topological phase of 2 in Fig.~\ref{fig:22}(b). The eigen-spectra over 200 chain configurations with disorders for each $\delta$ are shown in Fig.~\ref{fig:44}(a), where the red/gray dots mark the edge/bulk states. As expected, four topological edge states persist to appear in the bandgaps as if there were no disorder. Two edge profiles when $\delta=1$ mm are presented in Fig.~\ref{fig:44}(b), showing the localization of sound on the edges.   

This Letter provides an isospectral reduction approach to unearth the topology phenomena in the SSH3 model. An isospectrally- reduced model has been proposed, through which the bulk-edge correspondence of the SSH3 model has been established. An acoustic implementation has been conducted for verification. Our analysis has applications in the exploration of topological properties in various systems such as in photonics, plasmas, and mechanical systems.
  
\begin{acknowledgments}
This work is supported by the National Natural Science Foundation of China (Grant No.12204553) and Fundamental Research Funds for the Central Universities, Sun Yat-sen University (Grant No.24hytd008).
\end{acknowledgments}

\nocite{*}
\bibliography{rfs}

\begin{thebibliography}{49}%
\makeatletter
\providecommand \@ifxundefined [1]{%
 \@ifx{#1\undefined}
}%
\providecommand \@ifnum [1]{%
 \ifnum #1\expandafter \@firstoftwo
 \else \expandafter \@secondoftwo
 \fi
}%
\providecommand \@ifx [1]{%
 \ifx #1\expandafter \@firstoftwo
 \else \expandafter \@secondoftwo
 \fi
}%
\providecommand \natexlab [1]{#1}%
\providecommand \enquote  [1]{``#1''}%
\providecommand \bibnamefont  [1]{#1}%
\providecommand \bibfnamefont [1]{#1}%
\providecommand \citenamefont [1]{#1}%
\providecommand \href@noop [0]{\@secondoftwo}%
\providecommand \href [0]{\begingroup \@sanitize@url \@href}%
\providecommand \@href[1]{\@@startlink{#1}\@@href}%
\providecommand \@@href[1]{\endgroup#1\@@endlink}%
\providecommand \@sanitize@url [0]{\catcode `\\12\catcode `\$12\catcode
  `\&12\catcode `\#12\catcode `\^12\catcode `\_12\catcode `\%12\relax}%
\providecommand \@@startlink[1]{}%
\providecommand \@@endlink[0]{}%
\providecommand \url  [0]{\begingroup\@sanitize@url \@url }%
\providecommand \@url [1]{\endgroup\@href {#1}{\urlprefix }}%
\providecommand \urlprefix  [0]{URL }%
\providecommand \Eprint [0]{\href }%
\providecommand \doibase [0]{https://doi.org/}%
\providecommand \selectlanguage [0]{\@gobble}%
\providecommand \bibinfo  [0]{\@secondoftwo}%
\providecommand \bibfield  [0]{\@secondoftwo}%
\providecommand \translation [1]{[#1]}%
\providecommand \BibitemOpen [0]{}%
\providecommand \bibitemStop [0]{}%
\providecommand \bibitemNoStop [0]{.\EOS\space}%
\providecommand \EOS [0]{\spacefactor3000\relax}%
\providecommand \BibitemShut  [1]{\csname bibitem#1\endcsname}%
\let\auto@bib@innerbib\@empty
\bibitem [{\citenamefont {Yang}\ \emph {et~al.}(2015)\citenamefont {Yang},
  \citenamefont {Gao}, \citenamefont {Shi}, \citenamefont {Lin}, \citenamefont
  {Gao}, \citenamefont {Chong},\ and\ \citenamefont
  {Zhang}}]{PhysRevLett.114.114301}%
  \BibitemOpen
  \bibfield  {author} {\bibinfo {author} {\bibfnamefont {Z.}~\bibnamefont
  {Yang}}, \bibinfo {author} {\bibfnamefont {F.}~\bibnamefont {Gao}}, \bibinfo
  {author} {\bibfnamefont {X.}~\bibnamefont {Shi}}, \bibinfo {author}
  {\bibfnamefont {X.}~\bibnamefont {Lin}}, \bibinfo {author} {\bibfnamefont
  {Z.}~\bibnamefont {Gao}}, \bibinfo {author} {\bibfnamefont {Y.}~\bibnamefont
  {Chong}},\ and\ \bibinfo {author} {\bibfnamefont {B.}~\bibnamefont {Zhang}},\
  }\bibfield  {title} {\bibinfo {title} {Topological acoustics},\ }\href
  {https://doi.org/10.1103/PhysRevLett.114.114301} {\bibfield  {journal}
  {\bibinfo  {journal} {Phys. Rev. Lett.}\ }\textbf {\bibinfo {volume} {114}},\
  \bibinfo {pages} {114301} (\bibinfo {year} {2015})}\BibitemShut {NoStop}%
\bibitem [{\citenamefont {Ye}\ \emph {et~al.}(2022)\citenamefont {Ye},
  \citenamefont {Qiu}, \citenamefont {Xiao}, \citenamefont {Li}, \citenamefont
  {Du}, \citenamefont {Ke},\ and\ \citenamefont {Liu}}]{ye2022topological}%
  \BibitemOpen
  \bibfield  {author} {\bibinfo {author} {\bibfnamefont {L.}~\bibnamefont
  {Ye}}, \bibinfo {author} {\bibfnamefont {C.}~\bibnamefont {Qiu}}, \bibinfo
  {author} {\bibfnamefont {M.}~\bibnamefont {Xiao}}, \bibinfo {author}
  {\bibfnamefont {T.}~\bibnamefont {Li}}, \bibinfo {author} {\bibfnamefont
  {J.}~\bibnamefont {Du}}, \bibinfo {author} {\bibfnamefont {M.}~\bibnamefont
  {Ke}},\ and\ \bibinfo {author} {\bibfnamefont {Z.}~\bibnamefont {Liu}},\
  }\bibfield  {title} {\bibinfo {title} {Topological dislocation modes in
  three-dimensional acoustic topological insulators},\ }\href@noop {}
  {\bibfield  {journal} {\bibinfo  {journal} {Nature Communications}\ }\textbf
  {\bibinfo {volume} {13}},\ \bibinfo {pages} {508} (\bibinfo {year}
  {2022})}\BibitemShut {NoStop}%
\bibitem [{\citenamefont {Ozawa}\ \emph {et~al.}(2019)\citenamefont {Ozawa},
  \citenamefont {Price}, \citenamefont {Amo}, \citenamefont {Goldman},
  \citenamefont {Hafezi}, \citenamefont {Lu}, \citenamefont {Rechtsman},
  \citenamefont {Schuster}, \citenamefont {Simon}, \citenamefont {Zilberberg}
  \emph {et~al.}}]{ozawa2019topological}%
  \BibitemOpen
  \bibfield  {author} {\bibinfo {author} {\bibfnamefont {T.}~\bibnamefont
  {Ozawa}}, \bibinfo {author} {\bibfnamefont {H.~M.}\ \bibnamefont {Price}},
  \bibinfo {author} {\bibfnamefont {A.}~\bibnamefont {Amo}}, \bibinfo {author}
  {\bibfnamefont {N.}~\bibnamefont {Goldman}}, \bibinfo {author} {\bibfnamefont
  {M.}~\bibnamefont {Hafezi}}, \bibinfo {author} {\bibfnamefont
  {L.}~\bibnamefont {Lu}}, \bibinfo {author} {\bibfnamefont {M.~C.}\
  \bibnamefont {Rechtsman}}, \bibinfo {author} {\bibfnamefont {D.}~\bibnamefont
  {Schuster}}, \bibinfo {author} {\bibfnamefont {J.}~\bibnamefont {Simon}},
  \bibinfo {author} {\bibfnamefont {O.}~\bibnamefont {Zilberberg}}, \emph
  {et~al.},\ }\bibfield  {title} {\bibinfo {title} {Topological photonics},\
  }\href@noop {} {\bibfield  {journal} {\bibinfo  {journal} {Reviews of Modern
  Physics}\ }\textbf {\bibinfo {volume} {91}},\ \bibinfo {pages} {015006}
  (\bibinfo {year} {2019})}\BibitemShut {NoStop}%
\bibitem [{\citenamefont {Liao}\ \emph {et~al.}(2024)\citenamefont {Liao},
  \citenamefont {Zhang}, \citenamefont {Wang}, \citenamefont {Zhang},
  \citenamefont {Cortijo}, \citenamefont {Vozmediano}, \citenamefont {Guinea},
  \citenamefont {Cheng}, \citenamefont {Liu},\ and\ \citenamefont
  {Christensen}}]{liao2024visualizing}%
  \BibitemOpen
  \bibfield  {author} {\bibinfo {author} {\bibfnamefont {D.}~\bibnamefont
  {Liao}}, \bibinfo {author} {\bibfnamefont {J.}~\bibnamefont {Zhang}},
  \bibinfo {author} {\bibfnamefont {S.}~\bibnamefont {Wang}}, \bibinfo {author}
  {\bibfnamefont {Z.}~\bibnamefont {Zhang}}, \bibinfo {author} {\bibfnamefont
  {A.}~\bibnamefont {Cortijo}}, \bibinfo {author} {\bibfnamefont {M.~A.}\
  \bibnamefont {Vozmediano}}, \bibinfo {author} {\bibfnamefont
  {F.}~\bibnamefont {Guinea}}, \bibinfo {author} {\bibfnamefont
  {Y.}~\bibnamefont {Cheng}}, \bibinfo {author} {\bibfnamefont
  {X.}~\bibnamefont {Liu}},\ and\ \bibinfo {author} {\bibfnamefont
  {J.}~\bibnamefont {Christensen}},\ }\bibfield  {title} {\bibinfo {title}
  {Visualizing the topological pentagon states of a giant c540 metamaterial},\
  }\href@noop {} {\bibfield  {journal} {\bibinfo  {journal} {Nature
  Communications}\ }\textbf {\bibinfo {volume} {15}},\ \bibinfo {pages} {9644}
  (\bibinfo {year} {2024})}\BibitemShut {NoStop}%
\bibitem [{\citenamefont {Zhang}\ \emph {et~al.}(2023)\citenamefont {Zhang},
  \citenamefont {Zangeneh-Nejad}, \citenamefont {Chen}, \citenamefont {Lu},\
  and\ \citenamefont {Christensen}}]{zhang2023second}%
  \BibitemOpen
  \bibfield  {author} {\bibinfo {author} {\bibfnamefont {X.}~\bibnamefont
  {Zhang}}, \bibinfo {author} {\bibfnamefont {F.}~\bibnamefont
  {Zangeneh-Nejad}}, \bibinfo {author} {\bibfnamefont {Z.-G.}\ \bibnamefont
  {Chen}}, \bibinfo {author} {\bibfnamefont {M.-H.}\ \bibnamefont {Lu}},\ and\
  \bibinfo {author} {\bibfnamefont {J.}~\bibnamefont {Christensen}},\
  }\bibfield  {title} {\bibinfo {title} {A second wave of topological phenomena
  in photonics and acoustics},\ }\href@noop {} {\bibfield  {journal} {\bibinfo
  {journal} {Nature}\ }\textbf {\bibinfo {volume} {618}},\ \bibinfo {pages}
  {687} (\bibinfo {year} {2023})}\BibitemShut {NoStop}%
\bibitem [{\citenamefont {Zheng}\ \emph {et~al.}(2019)\citenamefont {Zheng},
  \citenamefont {Achilleos}, \citenamefont {Richoux}, \citenamefont
  {Theocharis},\ and\ \citenamefont {Pagneux}}]{PhysRevApplied.12.034014}%
  \BibitemOpen
  \bibfield  {author} {\bibinfo {author} {\bibfnamefont {L.-Y.}\ \bibnamefont
  {Zheng}}, \bibinfo {author} {\bibfnamefont {V.}~\bibnamefont {Achilleos}},
  \bibinfo {author} {\bibfnamefont {O.}~\bibnamefont {Richoux}}, \bibinfo
  {author} {\bibfnamefont {G.}~\bibnamefont {Theocharis}},\ and\ \bibinfo
  {author} {\bibfnamefont {V.}~\bibnamefont {Pagneux}},\ }\bibfield  {title}
  {\bibinfo {title} {Observation of edge waves in a two-dimensional
  su-schrieffer-heeger acoustic network},\ }\href
  {https://doi.org/10.1103/PhysRevApplied.12.034014} {\bibfield  {journal}
  {\bibinfo  {journal} {Phys. Rev. Appl.}\ }\textbf {\bibinfo {volume} {12}},\
  \bibinfo {pages} {034014} (\bibinfo {year} {2019})}\BibitemShut {NoStop}%
\bibitem [{\citenamefont {Zhang}\ \emph {et~al.}(2021)\citenamefont {Zhang},
  \citenamefont {Delplace},\ and\ \citenamefont {Fleury}}]{zhang2021superior}%
  \BibitemOpen
  \bibfield  {author} {\bibinfo {author} {\bibfnamefont {Z.}~\bibnamefont
  {Zhang}}, \bibinfo {author} {\bibfnamefont {P.}~\bibnamefont {Delplace}},\
  and\ \bibinfo {author} {\bibfnamefont {R.}~\bibnamefont {Fleury}},\
  }\bibfield  {title} {\bibinfo {title} {Superior robustness of anomalous
  non-reciprocal topological edge states},\ }\href@noop {} {\bibfield
  {journal} {\bibinfo  {journal} {Nature}\ }\textbf {\bibinfo {volume} {598}},\
  \bibinfo {pages} {293} (\bibinfo {year} {2021})}\BibitemShut {NoStop}%
\bibitem [{\citenamefont {Jin}\ \emph {et~al.}(2018)\citenamefont {Jin},
  \citenamefont {Torrent},\ and\ \citenamefont
  {Djafari-Rouhani}}]{jin2018robustness}%
  \BibitemOpen
  \bibfield  {author} {\bibinfo {author} {\bibfnamefont {Y.}~\bibnamefont
  {Jin}}, \bibinfo {author} {\bibfnamefont {D.}~\bibnamefont {Torrent}},\ and\
  \bibinfo {author} {\bibfnamefont {B.}~\bibnamefont {Djafari-Rouhani}},\
  }\bibfield  {title} {\bibinfo {title} {Robustness of conventional and
  topologically protected edge states in phononic crystal plates},\ }\href@noop
  {} {\bibfield  {journal} {\bibinfo  {journal} {Physical Review B}\ }\textbf
  {\bibinfo {volume} {98}},\ \bibinfo {pages} {054307} (\bibinfo {year}
  {2018})}\BibitemShut {NoStop}%
\bibitem [{\citenamefont {Wang}\ \emph {et~al.}(2021)\citenamefont {Wang},
  \citenamefont {Yang}, \citenamefont {Dai},\ and\ \citenamefont
  {Xu}}]{wang2021structural}%
  \BibitemOpen
  \bibfield  {author} {\bibinfo {author} {\bibfnamefont {J.-H.}\ \bibnamefont
  {Wang}}, \bibinfo {author} {\bibfnamefont {Y.-B.}\ \bibnamefont {Yang}},
  \bibinfo {author} {\bibfnamefont {N.}~\bibnamefont {Dai}},\ and\ \bibinfo
  {author} {\bibfnamefont {Y.}~\bibnamefont {Xu}},\ }\bibfield  {title}
  {\bibinfo {title} {Structural-disorder-induced second-order topological
  insulators in three dimensions},\ }\href@noop {} {\bibfield  {journal}
  {\bibinfo  {journal} {Physical Review Letters}\ }\textbf {\bibinfo {volume}
  {126}},\ \bibinfo {pages} {206404} (\bibinfo {year} {2021})}\BibitemShut
  {NoStop}%
\bibitem [{\citenamefont {Graf}\ and\ \citenamefont
  {Porta}(2013)}]{graf2013bulk}%
  \BibitemOpen
  \bibfield  {author} {\bibinfo {author} {\bibfnamefont {G.~M.}\ \bibnamefont
  {Graf}}\ and\ \bibinfo {author} {\bibfnamefont {M.}~\bibnamefont {Porta}},\
  }\bibfield  {title} {\bibinfo {title} {Bulk-edge correspondence for
  two-dimensional topological insulators},\ }\href@noop {} {\bibfield
  {journal} {\bibinfo  {journal} {Communications in Mathematical Physics}\
  }\textbf {\bibinfo {volume} {324}},\ \bibinfo {pages} {851} (\bibinfo {year}
  {2013})}\BibitemShut {NoStop}%
\bibitem [{\citenamefont {Kawarabayashi}\ and\ \citenamefont
  {Hatsugai}(2021)}]{kawarabayashi2021bulk}%
  \BibitemOpen
  \bibfield  {author} {\bibinfo {author} {\bibfnamefont {T.}~\bibnamefont
  {Kawarabayashi}}\ and\ \bibinfo {author} {\bibfnamefont {Y.}~\bibnamefont
  {Hatsugai}},\ }\bibfield  {title} {\bibinfo {title} {Bulk-edge correspondence
  with generalized chiral symmetry},\ }\href@noop {} {\bibfield  {journal}
  {\bibinfo  {journal} {Physical Review B}\ }\textbf {\bibinfo {volume}
  {103}},\ \bibinfo {pages} {205306} (\bibinfo {year} {2021})}\BibitemShut
  {NoStop}%
\bibitem [{\citenamefont {Mong}\ and\ \citenamefont
  {Shivamoggi}(2011)}]{mong2011edge}%
  \BibitemOpen
  \bibfield  {author} {\bibinfo {author} {\bibfnamefont {R.~S.}\ \bibnamefont
  {Mong}}\ and\ \bibinfo {author} {\bibfnamefont {V.}~\bibnamefont
  {Shivamoggi}},\ }\bibfield  {title} {\bibinfo {title} {Edge states and the
  bulk-boundary correspondence in dirac hamiltonians},\ }\href@noop {}
  {\bibfield  {journal} {\bibinfo  {journal} {Physical Review B—Condensed
  Matter and Materials Physics}\ }\textbf {\bibinfo {volume} {83}},\ \bibinfo
  {pages} {125109} (\bibinfo {year} {2011})}\BibitemShut {NoStop}%
\bibitem [{\citenamefont {Asb{\'o}th}\ \emph {et~al.}(2016)\citenamefont
  {Asb{\'o}th}, \citenamefont {Oroszl{\'a}ny},\ and\ \citenamefont
  {P{\'a}lyi}}]{asboth2016short}%
  \BibitemOpen
  \bibfield  {author} {\bibinfo {author} {\bibfnamefont {J.~K.}\ \bibnamefont
  {Asb{\'o}th}}, \bibinfo {author} {\bibfnamefont {L.}~\bibnamefont
  {Oroszl{\'a}ny}},\ and\ \bibinfo {author} {\bibfnamefont {A.}~\bibnamefont
  {P{\'a}lyi}},\ }\bibfield  {title} {\bibinfo {title} {A short course on
  topological insulators},\ }\href@noop {} {\bibfield  {journal} {\bibinfo
  {journal} {Lecture notes in physics}\ }\textbf {\bibinfo {volume} {919}}
  (\bibinfo {year} {2016})}\BibitemShut {NoStop}%
\bibitem [{\citenamefont {Thouless}\ \emph {et~al.}(1982)\citenamefont
  {Thouless}, \citenamefont {Kohmoto}, \citenamefont {Nightingale},\ and\
  \citenamefont {den Nijs}}]{thouless1982quantized}%
  \BibitemOpen
  \bibfield  {author} {\bibinfo {author} {\bibfnamefont {D.~J.}\ \bibnamefont
  {Thouless}}, \bibinfo {author} {\bibfnamefont {M.}~\bibnamefont {Kohmoto}},
  \bibinfo {author} {\bibfnamefont {M.~P.}\ \bibnamefont {Nightingale}},\ and\
  \bibinfo {author} {\bibfnamefont {M.}~\bibnamefont {den Nijs}},\ }\bibfield
  {title} {\bibinfo {title} {Quantized hall conductance in a two-dimensional
  periodic potential},\ }\href@noop {} {\bibfield  {journal} {\bibinfo
  {journal} {Physical review letters}\ }\textbf {\bibinfo {volume} {49}},\
  \bibinfo {pages} {405} (\bibinfo {year} {1982})}\BibitemShut {NoStop}%
\bibitem [{\citenamefont {Haldane}(1988)}]{haldane1988model}%
  \BibitemOpen
  \bibfield  {author} {\bibinfo {author} {\bibfnamefont {F.~D.~M.}\
  \bibnamefont {Haldane}},\ }\bibfield  {title} {\bibinfo {title} {Model for a
  quantum hall effect without landau levels: Condensed-matter realization of
  the" parity anomaly"},\ }\href@noop {} {\bibfield  {journal} {\bibinfo
  {journal} {Physical review letters}\ }\textbf {\bibinfo {volume} {61}},\
  \bibinfo {pages} {2015} (\bibinfo {year} {1988})}\BibitemShut {NoStop}%
\bibitem [{\citenamefont {Xue}\ \emph {et~al.}(2022)\citenamefont {Xue},
  \citenamefont {Yang},\ and\ \citenamefont {Zhang}}]{xue2022topological}%
  \BibitemOpen
  \bibfield  {author} {\bibinfo {author} {\bibfnamefont {H.}~\bibnamefont
  {Xue}}, \bibinfo {author} {\bibfnamefont {Y.}~\bibnamefont {Yang}},\ and\
  \bibinfo {author} {\bibfnamefont {B.}~\bibnamefont {Zhang}},\ }\bibfield
  {title} {\bibinfo {title} {Topological acoustics},\ }\href@noop {} {\bibfield
   {journal} {\bibinfo  {journal} {Nature Reviews Materials}\ }\textbf
  {\bibinfo {volume} {7}},\ \bibinfo {pages} {974} (\bibinfo {year}
  {2022})}\BibitemShut {NoStop}%
\bibitem [{\citenamefont {Kane}\ and\ \citenamefont
  {Mele}(2005)}]{kane2005quantum}%
  \BibitemOpen
  \bibfield  {author} {\bibinfo {author} {\bibfnamefont {C.~L.}\ \bibnamefont
  {Kane}}\ and\ \bibinfo {author} {\bibfnamefont {E.~J.}\ \bibnamefont
  {Mele}},\ }\bibfield  {title} {\bibinfo {title} {Quantum spin hall effect in
  graphene},\ }\href@noop {} {\bibfield  {journal} {\bibinfo  {journal}
  {Physical review letters}\ }\textbf {\bibinfo {volume} {95}},\ \bibinfo
  {pages} {226801} (\bibinfo {year} {2005})}\BibitemShut {NoStop}%
\bibitem [{\citenamefont {S{\"u}sstrunk}\ and\ \citenamefont
  {Huber}(2015)}]{susstrunk2015observation}%
  \BibitemOpen
  \bibfield  {author} {\bibinfo {author} {\bibfnamefont {R.}~\bibnamefont
  {S{\"u}sstrunk}}\ and\ \bibinfo {author} {\bibfnamefont {S.~D.}\ \bibnamefont
  {Huber}},\ }\bibfield  {title} {\bibinfo {title} {Observation of phononic
  helical edge states in a mechanical topological insulator},\ }\href@noop {}
  {\bibfield  {journal} {\bibinfo  {journal} {Science}\ }\textbf {\bibinfo
  {volume} {349}},\ \bibinfo {pages} {47} (\bibinfo {year} {2015})}\BibitemShut
  {NoStop}%
\bibitem [{\citenamefont {He}\ \emph {et~al.}(2016)\citenamefont {He},
  \citenamefont {Ni}, \citenamefont {Ge}, \citenamefont {Sun}, \citenamefont
  {Chen}, \citenamefont {Lu}, \citenamefont {Liu},\ and\ \citenamefont
  {Chen}}]{he2016acoustic}%
  \BibitemOpen
  \bibfield  {author} {\bibinfo {author} {\bibfnamefont {C.}~\bibnamefont
  {He}}, \bibinfo {author} {\bibfnamefont {X.}~\bibnamefont {Ni}}, \bibinfo
  {author} {\bibfnamefont {H.}~\bibnamefont {Ge}}, \bibinfo {author}
  {\bibfnamefont {X.-C.}\ \bibnamefont {Sun}}, \bibinfo {author} {\bibfnamefont
  {Y.-B.}\ \bibnamefont {Chen}}, \bibinfo {author} {\bibfnamefont {M.-H.}\
  \bibnamefont {Lu}}, \bibinfo {author} {\bibfnamefont {X.-P.}\ \bibnamefont
  {Liu}},\ and\ \bibinfo {author} {\bibfnamefont {Y.-F.}\ \bibnamefont
  {Chen}},\ }\bibfield  {title} {\bibinfo {title} {Acoustic topological
  insulator and robust one-way sound transport},\ }\href@noop {} {\bibfield
  {journal} {\bibinfo  {journal} {Nature physics}\ }\textbf {\bibinfo {volume}
  {12}},\ \bibinfo {pages} {1124} (\bibinfo {year} {2016})}\BibitemShut
  {NoStop}%
\bibitem [{\citenamefont {Song}\ and\ \citenamefont
  {Prodan}(2015)}]{song2015quantization}%
  \BibitemOpen
  \bibfield  {author} {\bibinfo {author} {\bibfnamefont {J.}~\bibnamefont
  {Song}}\ and\ \bibinfo {author} {\bibfnamefont {E.}~\bibnamefont {Prodan}},\
  }\bibfield  {title} {\bibinfo {title} {Quantization of topological invariants
  under symmetry-breaking disorder},\ }\href@noop {} {\bibfield  {journal}
  {\bibinfo  {journal} {Physical Review B}\ }\textbf {\bibinfo {volume} {92}},\
  \bibinfo {pages} {195119} (\bibinfo {year} {2015})}\BibitemShut {NoStop}%
\bibitem [{\citenamefont {Gong}\ \emph {et~al.}(2018)\citenamefont {Gong},
  \citenamefont {Ashida}, \citenamefont {Kawabata}, \citenamefont {Takasan},
  \citenamefont {Higashikawa},\ and\ \citenamefont
  {Ueda}}]{gong2018topological}%
  \BibitemOpen
  \bibfield  {author} {\bibinfo {author} {\bibfnamefont {Z.}~\bibnamefont
  {Gong}}, \bibinfo {author} {\bibfnamefont {Y.}~\bibnamefont {Ashida}},
  \bibinfo {author} {\bibfnamefont {K.}~\bibnamefont {Kawabata}}, \bibinfo
  {author} {\bibfnamefont {K.}~\bibnamefont {Takasan}}, \bibinfo {author}
  {\bibfnamefont {S.}~\bibnamefont {Higashikawa}},\ and\ \bibinfo {author}
  {\bibfnamefont {M.}~\bibnamefont {Ueda}},\ }\bibfield  {title} {\bibinfo
  {title} {Topological phases of non-hermitian systems},\ }\href@noop {}
  {\bibfield  {journal} {\bibinfo  {journal} {Physical Review X}\ }\textbf
  {\bibinfo {volume} {8}},\ \bibinfo {pages} {031079} (\bibinfo {year}
  {2018})}\BibitemShut {NoStop}%
\bibitem [{\citenamefont {Okuma}\ and\ \citenamefont
  {Sato}(2023)}]{okuma2023non}%
  \BibitemOpen
  \bibfield  {author} {\bibinfo {author} {\bibfnamefont {N.}~\bibnamefont
  {Okuma}}\ and\ \bibinfo {author} {\bibfnamefont {M.}~\bibnamefont {Sato}},\
  }\bibfield  {title} {\bibinfo {title} {Non-hermitian topological phenomena: A
  review},\ }\href@noop {} {\bibfield  {journal} {\bibinfo  {journal} {Annual
  Review of Condensed Matter Physics}\ }\textbf {\bibinfo {volume} {14}},\
  \bibinfo {pages} {83} (\bibinfo {year} {2023})}\BibitemShut {NoStop}%
\bibitem [{\citenamefont {Lieu}(2018)}]{lieu2018topological}%
  \BibitemOpen
  \bibfield  {author} {\bibinfo {author} {\bibfnamefont {S.}~\bibnamefont
  {Lieu}},\ }\bibfield  {title} {\bibinfo {title} {Topological phases in the
  non-hermitian su-schrieffer-heeger model},\ }\href@noop {} {\bibfield
  {journal} {\bibinfo  {journal} {Physical Review B}\ }\textbf {\bibinfo
  {volume} {97}},\ \bibinfo {pages} {045106} (\bibinfo {year}
  {2018})}\BibitemShut {NoStop}%
\bibitem [{\citenamefont {Zangeneh-Nejad}\ and\ \citenamefont
  {Fleury}(2019)}]{zangeneh2019nonlinear}%
  \BibitemOpen
  \bibfield  {author} {\bibinfo {author} {\bibfnamefont {F.}~\bibnamefont
  {Zangeneh-Nejad}}\ and\ \bibinfo {author} {\bibfnamefont {R.}~\bibnamefont
  {Fleury}},\ }\bibfield  {title} {\bibinfo {title} {Nonlinear second-order
  topological insulators},\ }\href@noop {} {\bibfield  {journal} {\bibinfo
  {journal} {Physical review letters}\ }\textbf {\bibinfo {volume} {123}},\
  \bibinfo {pages} {053902} (\bibinfo {year} {2019})}\BibitemShut {NoStop}%
\bibitem [{\citenamefont {Tuloup}\ \emph {et~al.}(2020)\citenamefont {Tuloup},
  \citenamefont {Bomantara}, \citenamefont {Lee},\ and\ \citenamefont
  {Gong}}]{tuloup2020nonlinearity}%
  \BibitemOpen
  \bibfield  {author} {\bibinfo {author} {\bibfnamefont {T.}~\bibnamefont
  {Tuloup}}, \bibinfo {author} {\bibfnamefont {R.~W.}\ \bibnamefont
  {Bomantara}}, \bibinfo {author} {\bibfnamefont {C.~H.}\ \bibnamefont {Lee}},\
  and\ \bibinfo {author} {\bibfnamefont {J.}~\bibnamefont {Gong}},\ }\bibfield
  {title} {\bibinfo {title} {Nonlinearity induced topological physics in
  momentum space and real space},\ }\href@noop {} {\bibfield  {journal}
  {\bibinfo  {journal} {Physical Review B}\ }\textbf {\bibinfo {volume}
  {102}},\ \bibinfo {pages} {115411} (\bibinfo {year} {2020})}\BibitemShut
  {NoStop}%
\bibitem [{\citenamefont {Jiang}\ \emph {et~al.}(2021)\citenamefont {Jiang},
  \citenamefont {Bouhon}, \citenamefont {Lin}, \citenamefont {Zhou},
  \citenamefont {Hou}, \citenamefont {Li}, \citenamefont {Slager},\ and\
  \citenamefont {Jiang}}]{jiang2021experimental}%
  \BibitemOpen
  \bibfield  {author} {\bibinfo {author} {\bibfnamefont {B.}~\bibnamefont
  {Jiang}}, \bibinfo {author} {\bibfnamefont {A.}~\bibnamefont {Bouhon}},
  \bibinfo {author} {\bibfnamefont {Z.-K.}\ \bibnamefont {Lin}}, \bibinfo
  {author} {\bibfnamefont {X.}~\bibnamefont {Zhou}}, \bibinfo {author}
  {\bibfnamefont {B.}~\bibnamefont {Hou}}, \bibinfo {author} {\bibfnamefont
  {F.}~\bibnamefont {Li}}, \bibinfo {author} {\bibfnamefont {R.-J.}\
  \bibnamefont {Slager}},\ and\ \bibinfo {author} {\bibfnamefont {J.-H.}\
  \bibnamefont {Jiang}},\ }\bibfield  {title} {\bibinfo {title} {Experimental
  observation of non-abelian topological acoustic semimetals and their phase
  transitions},\ }\href@noop {} {\bibfield  {journal} {\bibinfo  {journal}
  {Nature Physics}\ }\textbf {\bibinfo {volume} {17}},\ \bibinfo {pages} {1239}
  (\bibinfo {year} {2021})}\BibitemShut {NoStop}%
\bibitem [{\citenamefont {Guo}\ \emph {et~al.}(2021)\citenamefont {Guo},
  \citenamefont {Jiang}, \citenamefont {Zhang}, \citenamefont {Zhang},
  \citenamefont {Zhang}, \citenamefont {Yang}, \citenamefont {Zhang},\ and\
  \citenamefont {Chan}}]{guo2021experimental}%
  \BibitemOpen
  \bibfield  {author} {\bibinfo {author} {\bibfnamefont {Q.}~\bibnamefont
  {Guo}}, \bibinfo {author} {\bibfnamefont {T.}~\bibnamefont {Jiang}}, \bibinfo
  {author} {\bibfnamefont {R.-Y.}\ \bibnamefont {Zhang}}, \bibinfo {author}
  {\bibfnamefont {L.}~\bibnamefont {Zhang}}, \bibinfo {author} {\bibfnamefont
  {Z.-Q.}\ \bibnamefont {Zhang}}, \bibinfo {author} {\bibfnamefont
  {B.}~\bibnamefont {Yang}}, \bibinfo {author} {\bibfnamefont {S.}~\bibnamefont
  {Zhang}},\ and\ \bibinfo {author} {\bibfnamefont {C.~T.}\ \bibnamefont
  {Chan}},\ }\bibfield  {title} {\bibinfo {title} {Experimental observation of
  non-abelian topological charges and edge states},\ }\href@noop {} {\bibfield
  {journal} {\bibinfo  {journal} {Nature}\ }\textbf {\bibinfo {volume} {594}},\
  \bibinfo {pages} {195} (\bibinfo {year} {2021})}\BibitemShut {NoStop}%
\bibitem [{\citenamefont {Martinez~Alvarez}\ and\ \citenamefont
  {Coutinho-Filho}(2019)}]{PhysRevA.99.013833}%
  \BibitemOpen
  \bibfield  {author} {\bibinfo {author} {\bibfnamefont {V.~M.}\ \bibnamefont
  {Martinez~Alvarez}}\ and\ \bibinfo {author} {\bibfnamefont {M.~D.}\
  \bibnamefont {Coutinho-Filho}},\ }\bibfield  {title} {\bibinfo {title} {Edge
  states in trimer lattices},\ }\href
  {https://doi.org/10.1103/PhysRevA.99.013833} {\bibfield  {journal} {\bibinfo
  {journal} {Phys. Rev. A}\ }\textbf {\bibinfo {volume} {99}},\ \bibinfo
  {pages} {013833} (\bibinfo {year} {2019})}\BibitemShut {NoStop}%
\bibitem [{\citenamefont {Anastasiadis}\ \emph {et~al.}(2022)\citenamefont
  {Anastasiadis}, \citenamefont {Styliaris}, \citenamefont {Chaunsali},
  \citenamefont {Theocharis},\ and\ \citenamefont
  {Diakonos}}]{Anastasiadis_2022}%
  \BibitemOpen
  \bibfield  {author} {\bibinfo {author} {\bibfnamefont {A.}~\bibnamefont
  {Anastasiadis}}, \bibinfo {author} {\bibfnamefont {G.}~\bibnamefont
  {Styliaris}}, \bibinfo {author} {\bibfnamefont {R.}~\bibnamefont
  {Chaunsali}}, \bibinfo {author} {\bibfnamefont {G.}~\bibnamefont
  {Theocharis}},\ and\ \bibinfo {author} {\bibfnamefont {F.~K.}\ \bibnamefont
  {Diakonos}},\ }\bibfield  {title} {\bibinfo {title} {Bulk-edge correspondence
  in the trimer su-schrieffer-heeger model},\ }\bibfield  {journal} {\bibinfo
  {journal} {Physical Review B}\ }\textbf {\bibinfo {volume} {106}},\ \href
  {https://doi.org/10.1103/physrevb.106.085109} {10.1103/physrevb.106.085109}
  (\bibinfo {year} {2022})\BibitemShut {NoStop}%
\bibitem [{\citenamefont {Eiles}\ \emph {et~al.}(2024)\citenamefont {Eiles},
  \citenamefont {W{\"a}chtler}, \citenamefont {Eisfeld},\ and\ \citenamefont
  {Rost}}]{eiles2024topological}%
  \BibitemOpen
  \bibfield  {author} {\bibinfo {author} {\bibfnamefont {M.~T.}\ \bibnamefont
  {Eiles}}, \bibinfo {author} {\bibfnamefont {C.~W.}\ \bibnamefont
  {W{\"a}chtler}}, \bibinfo {author} {\bibfnamefont {A.}~\bibnamefont
  {Eisfeld}},\ and\ \bibinfo {author} {\bibfnamefont {J.~M.}\ \bibnamefont
  {Rost}},\ }\bibfield  {title} {\bibinfo {title} {Topological edge states in a
  rydberg composite},\ }\href@noop {} {\bibfield  {journal} {\bibinfo
  {journal} {Physical Review B}\ }\textbf {\bibinfo {volume} {109}},\ \bibinfo
  {pages} {075422} (\bibinfo {year} {2024})}\BibitemShut {NoStop}%
\bibitem [{\citenamefont {Xie}\ \emph {et~al.}(2019)\citenamefont {Xie},
  \citenamefont {Gou}, \citenamefont {Xiao}, \citenamefont {Gadway},\ and\
  \citenamefont {Yan}}]{Xie_2019}%
  \BibitemOpen
  \bibfield  {author} {\bibinfo {author} {\bibfnamefont {D.}~\bibnamefont
  {Xie}}, \bibinfo {author} {\bibfnamefont {W.}~\bibnamefont {Gou}}, \bibinfo
  {author} {\bibfnamefont {T.}~\bibnamefont {Xiao}}, \bibinfo {author}
  {\bibfnamefont {B.}~\bibnamefont {Gadway}},\ and\ \bibinfo {author}
  {\bibfnamefont {B.}~\bibnamefont {Yan}},\ }\bibfield  {title} {\bibinfo
  {title} {Topological characterizations of an extended
  su–schrieffer–heeger model},\ }\bibfield  {journal} {\bibinfo  {journal}
  {npj Quantum Information}\ }\textbf {\bibinfo {volume} {5}},\ \href
  {https://doi.org/10.1038/s41534-019-0159-6} {10.1038/s41534-019-0159-6}
  (\bibinfo {year} {2019})\BibitemShut {NoStop}%
\bibitem [{\citenamefont {Chiu}\ \emph {et~al.}(2016)\citenamefont {Chiu},
  \citenamefont {Teo}, \citenamefont {Schnyder},\ and\ \citenamefont
  {Ryu}}]{chiu2016classification}%
  \BibitemOpen
  \bibfield  {author} {\bibinfo {author} {\bibfnamefont {C.-K.}\ \bibnamefont
  {Chiu}}, \bibinfo {author} {\bibfnamefont {J.~C.}\ \bibnamefont {Teo}},
  \bibinfo {author} {\bibfnamefont {A.~P.}\ \bibnamefont {Schnyder}},\ and\
  \bibinfo {author} {\bibfnamefont {S.}~\bibnamefont {Ryu}},\ }\bibfield
  {title} {\bibinfo {title} {Classification of topological quantum matter with
  symmetries},\ }\href@noop {} {\bibfield  {journal} {\bibinfo  {journal}
  {Reviews of Modern Physics}\ }\textbf {\bibinfo {volume} {88}},\ \bibinfo
  {pages} {035005} (\bibinfo {year} {2016})}\BibitemShut {NoStop}%
\bibitem [{\citenamefont {Wang}\ and\ \citenamefont
  {Gu}(2020)}]{wang2020construction}%
  \BibitemOpen
  \bibfield  {author} {\bibinfo {author} {\bibfnamefont {Q.-R.}\ \bibnamefont
  {Wang}}\ and\ \bibinfo {author} {\bibfnamefont {Z.-C.}\ \bibnamefont {Gu}},\
  }\bibfield  {title} {\bibinfo {title} {Construction and classification of
  symmetry-protected topological phases in interacting fermion systems},\
  }\href@noop {} {\bibfield  {journal} {\bibinfo  {journal} {Physical Review
  X}\ }\textbf {\bibinfo {volume} {10}},\ \bibinfo {pages} {031055} (\bibinfo
  {year} {2020})}\BibitemShut {NoStop}%
\bibitem [{\citenamefont {Sougleridis}\ \emph {et~al.}(2024)\citenamefont
  {Sougleridis}, \citenamefont {Anastasiadis}, \citenamefont {Richoux},
  \citenamefont {Achilleos}, \citenamefont {Theocharis}, \citenamefont
  {Pagneux},\ and\ \citenamefont {Diakonos}}]{sougleridis2024existence}%
  \BibitemOpen
  \bibfield  {author} {\bibinfo {author} {\bibfnamefont {I.~I.}\ \bibnamefont
  {Sougleridis}}, \bibinfo {author} {\bibfnamefont {A.}~\bibnamefont
  {Anastasiadis}}, \bibinfo {author} {\bibfnamefont {O.}~\bibnamefont
  {Richoux}}, \bibinfo {author} {\bibfnamefont {V.}~\bibnamefont {Achilleos}},
  \bibinfo {author} {\bibfnamefont {G.}~\bibnamefont {Theocharis}}, \bibinfo
  {author} {\bibfnamefont {V.}~\bibnamefont {Pagneux}},\ and\ \bibinfo {author}
  {\bibfnamefont {F.}~\bibnamefont {Diakonos}},\ }\bibfield  {title} {\bibinfo
  {title} {Existence and characterization of edge states in an acoustic trimer
  su-schrieffer-heeger model},\ }\href@noop {} {\bibfield  {journal} {\bibinfo
  {journal} {arXiv preprint arXiv:2401.14264}\ } (\bibinfo {year}
  {2024})}\BibitemShut {NoStop}%
\bibitem [{\citenamefont {Kempton}\ \emph {et~al.}(2020)\citenamefont
  {Kempton}, \citenamefont {Sinkovic}, \citenamefont {Smith},\ and\
  \citenamefont {Webb}}]{kempton2020characterizing}%
  \BibitemOpen
  \bibfield  {author} {\bibinfo {author} {\bibfnamefont {M.}~\bibnamefont
  {Kempton}}, \bibinfo {author} {\bibfnamefont {J.}~\bibnamefont {Sinkovic}},
  \bibinfo {author} {\bibfnamefont {D.}~\bibnamefont {Smith}},\ and\ \bibinfo
  {author} {\bibfnamefont {B.}~\bibnamefont {Webb}},\ }\bibfield  {title}
  {\bibinfo {title} {Characterizing cospectral vertices via isospectral
  reduction},\ }\href@noop {} {\bibfield  {journal} {\bibinfo  {journal}
  {Linear Algebra and its Applications}\ }\textbf {\bibinfo {volume} {594}},\
  \bibinfo {pages} {226} (\bibinfo {year} {2020})}\BibitemShut {NoStop}%
\bibitem [{\citenamefont {Smith}\ and\ \citenamefont
  {Webb}(2019)}]{smith2019hidden}%
  \BibitemOpen
  \bibfield  {author} {\bibinfo {author} {\bibfnamefont {D.}~\bibnamefont
  {Smith}}\ and\ \bibinfo {author} {\bibfnamefont {B.}~\bibnamefont {Webb}},\
  }\bibfield  {title} {\bibinfo {title} {Hidden symmetries in real and
  theoretical networks},\ }\href@noop {} {\bibfield  {journal} {\bibinfo
  {journal} {Physica A: Statistical Mechanics and its Applications}\ }\textbf
  {\bibinfo {volume} {514}},\ \bibinfo {pages} {855} (\bibinfo {year}
  {2019})}\BibitemShut {NoStop}%
\bibitem [{\citenamefont {Bunimovich}\ and\ \citenamefont
  {Webb}(2011)}]{bunimovich2011isospectral}%
  \BibitemOpen
  \bibfield  {author} {\bibinfo {author} {\bibfnamefont {L.}~\bibnamefont
  {Bunimovich}}\ and\ \bibinfo {author} {\bibfnamefont {B.}~\bibnamefont
  {Webb}},\ }\bibfield  {title} {\bibinfo {title} {Isospectral graph
  transformations, spectral equivalence, and global stability of dynamical
  networks},\ }\href@noop {} {\bibfield  {journal} {\bibinfo  {journal}
  {Nonlinearity}\ }\textbf {\bibinfo {volume} {25}},\ \bibinfo {pages} {211}
  (\bibinfo {year} {2011})}\BibitemShut {NoStop}%
\bibitem [{\citenamefont {R{\"o}ntgen}\ \emph {et~al.}(2023)\citenamefont
  {R{\"o}ntgen}, \citenamefont {Morfonios}, \citenamefont {Schmelcher},\ and\
  \citenamefont {Pagneux}}]{rontgen2023hidden}%
  \BibitemOpen
  \bibfield  {author} {\bibinfo {author} {\bibfnamefont {M.}~\bibnamefont
  {R{\"o}ntgen}}, \bibinfo {author} {\bibfnamefont {C.~V.}\ \bibnamefont
  {Morfonios}}, \bibinfo {author} {\bibfnamefont {P.}~\bibnamefont
  {Schmelcher}},\ and\ \bibinfo {author} {\bibfnamefont {V.}~\bibnamefont
  {Pagneux}},\ }\bibfield  {title} {\bibinfo {title} {Hidden symmetries in
  acoustic wave systems},\ }\href@noop {} {\bibfield  {journal} {\bibinfo
  {journal} {Physical Review Letters}\ }\textbf {\bibinfo {volume} {130}},\
  \bibinfo {pages} {077201} (\bibinfo {year} {2023})}\BibitemShut {NoStop}%
\bibitem [{\citenamefont {Bunimovich}\ and\ \citenamefont
  {Webb}(2014)}]{bunimovich2014improved}%
  \BibitemOpen
  \bibfield  {author} {\bibinfo {author} {\bibfnamefont {L.}~\bibnamefont
  {Bunimovich}}\ and\ \bibinfo {author} {\bibfnamefont {B.}~\bibnamefont
  {Webb}},\ }\bibfield  {title} {\bibinfo {title} {Improved estimates of
  survival probabilities via isospectral transformations},\ }in\ \href@noop {}
  {\emph {\bibinfo {booktitle} {Ergodic Theory, Open Dynamics, and Coherent
  Structures}}}\ (\bibinfo {organization} {Springer},\ \bibinfo {year} {2014})\
  pp.\ \bibinfo {pages} {119--135}\BibitemShut {NoStop}%
\bibitem [{\citenamefont {Hou}\ and\ \citenamefont
  {Chen}(2018)}]{hou2018hidden}%
  \BibitemOpen
  \bibfield  {author} {\bibinfo {author} {\bibfnamefont {J.-M.}\ \bibnamefont
  {Hou}}\ and\ \bibinfo {author} {\bibfnamefont {W.}~\bibnamefont {Chen}},\
  }\bibfield  {title} {\bibinfo {title} {Hidden antiunitary symmetry behind
  “accidental” degeneracy and its protection of degeneracy},\ }\href@noop
  {} {\bibfield  {journal} {\bibinfo  {journal} {Frontiers of Physics}\
  }\textbf {\bibinfo {volume} {13}},\ \bibinfo {pages} {1} (\bibinfo {year}
  {2018})}\BibitemShut {NoStop}%
\bibitem [{\citenamefont {Li}\ \emph {et~al.}(2024)\citenamefont {Li},
  \citenamefont {Zhang}, \citenamefont {Liu},\ and\ \citenamefont
  {Liu}}]{li2024group}%
  \BibitemOpen
  \bibfield  {author} {\bibinfo {author} {\bibfnamefont {J.}~\bibnamefont
  {Li}}, \bibinfo {author} {\bibfnamefont {A.}~\bibnamefont {Zhang}}, \bibinfo
  {author} {\bibfnamefont {Y.}~\bibnamefont {Liu}},\ and\ \bibinfo {author}
  {\bibfnamefont {Q.}~\bibnamefont {Liu}},\ }\bibfield  {title} {\bibinfo
  {title} {Group theory on quasisymmetry and protected near degeneracy},\
  }\href@noop {} {\bibfield  {journal} {\bibinfo  {journal} {Physical Review
  Letters}\ }\textbf {\bibinfo {volume} {133}},\ \bibinfo {pages} {026402}
  (\bibinfo {year} {2024})}\BibitemShut {NoStop}%
\bibitem [{\citenamefont {R{\"o}ntgen}\ \emph {et~al.}(2021)\citenamefont
  {R{\"o}ntgen}, \citenamefont {Pyzh}, \citenamefont {Morfonios}, \citenamefont
  {Palaiodimopoulos}, \citenamefont {Diakonos},\ and\ \citenamefont
  {Schmelcher}}]{rontgen2021latent}%
  \BibitemOpen
  \bibfield  {author} {\bibinfo {author} {\bibfnamefont {M.}~\bibnamefont
  {R{\"o}ntgen}}, \bibinfo {author} {\bibfnamefont {M.}~\bibnamefont {Pyzh}},
  \bibinfo {author} {\bibfnamefont {C.}~\bibnamefont {Morfonios}}, \bibinfo
  {author} {\bibfnamefont {N.}~\bibnamefont {Palaiodimopoulos}}, \bibinfo
  {author} {\bibfnamefont {F.}~\bibnamefont {Diakonos}},\ and\ \bibinfo
  {author} {\bibfnamefont {P.}~\bibnamefont {Schmelcher}},\ }\bibfield  {title}
  {\bibinfo {title} {Latent symmetry induced degeneracies},\ }\href@noop {}
  {\bibfield  {journal} {\bibinfo  {journal} {Physical Review Letters}\
  }\textbf {\bibinfo {volume} {126}},\ \bibinfo {pages} {180601} (\bibinfo
  {year} {2021})}\BibitemShut {NoStop}%
\bibitem [{\citenamefont {R{\"o}ntgen}\ \emph {et~al.}(2024)\citenamefont
  {R{\"o}ntgen}, \citenamefont {Chen}, \citenamefont {Gao}, \citenamefont
  {Pyzh}, \citenamefont {Schmelcher}, \citenamefont {Pagneux}, \citenamefont
  {Achilleos},\ and\ \citenamefont {Coutant}}]{rontgen2024topological}%
  \BibitemOpen
  \bibfield  {author} {\bibinfo {author} {\bibfnamefont {M.}~\bibnamefont
  {R{\"o}ntgen}}, \bibinfo {author} {\bibfnamefont {X.}~\bibnamefont {Chen}},
  \bibinfo {author} {\bibfnamefont {W.}~\bibnamefont {Gao}}, \bibinfo {author}
  {\bibfnamefont {M.}~\bibnamefont {Pyzh}}, \bibinfo {author} {\bibfnamefont
  {P.}~\bibnamefont {Schmelcher}}, \bibinfo {author} {\bibfnamefont
  {V.}~\bibnamefont {Pagneux}}, \bibinfo {author} {\bibfnamefont
  {V.}~\bibnamefont {Achilleos}},\ and\ \bibinfo {author} {\bibfnamefont
  {A.}~\bibnamefont {Coutant}},\ }\bibfield  {title} {\bibinfo {title}
  {Topological states protected by hidden symmetry},\ }\href@noop {} {\bibfield
   {journal} {\bibinfo  {journal} {Physical Review B}\ }\textbf {\bibinfo
  {volume} {110}},\ \bibinfo {pages} {035106} (\bibinfo {year}
  {2024})}\BibitemShut {NoStop}%
\bibitem [{\citenamefont {Zheng}\ \emph {et~al.}(2023)\citenamefont {Zheng},
  \citenamefont {Li}, \citenamefont {Zhang},\ and\ \citenamefont
  {Huang}}]{zheng2023robust}%
  \BibitemOpen
  \bibfield  {author} {\bibinfo {author} {\bibfnamefont {L.-Y.}\ \bibnamefont
  {Zheng}}, \bibinfo {author} {\bibfnamefont {Y.-F.}\ \bibnamefont {Li}},
  \bibinfo {author} {\bibfnamefont {J.}~\bibnamefont {Zhang}},\ and\ \bibinfo
  {author} {\bibfnamefont {Y.}~\bibnamefont {Huang}},\ }\bibfield  {title}
  {\bibinfo {title} {Robust topological edge states induced by latent mirror
  symmetry},\ }\href@noop {} {\bibfield  {journal} {\bibinfo  {journal}
  {Physical Review B}\ }\textbf {\bibinfo {volume} {108}},\ \bibinfo {pages}
  {L220303} (\bibinfo {year} {2023})}\BibitemShut {NoStop}%
\bibitem [{\citenamefont {S{\'a}}\ \emph {et~al.}(2023)\citenamefont {S{\'a}},
  \citenamefont {Ribeiro},\ and\ \citenamefont {Prosen}}]{sa2023symmetry}%
  \BibitemOpen
  \bibfield  {author} {\bibinfo {author} {\bibfnamefont {L.}~\bibnamefont
  {S{\'a}}}, \bibinfo {author} {\bibfnamefont {P.}~\bibnamefont {Ribeiro}},\
  and\ \bibinfo {author} {\bibfnamefont {T.}~\bibnamefont {Prosen}},\
  }\bibfield  {title} {\bibinfo {title} {Symmetry classification of many-body
  lindbladians: Tenfold way and beyond},\ }\href@noop {} {\bibfield  {journal}
  {\bibinfo  {journal} {Physical Review X}\ }\textbf {\bibinfo {volume} {13}},\
  \bibinfo {pages} {031019} (\bibinfo {year} {2023})}\BibitemShut {NoStop}%
\bibitem [{\citenamefont {Chen}\ \emph {et~al.}(2019)\citenamefont {Chen},
  \citenamefont {Kapustin}, \citenamefont {Turzillo},\ and\ \citenamefont
  {You}}]{chen2019free}%
  \BibitemOpen
  \bibfield  {author} {\bibinfo {author} {\bibfnamefont {Y.-A.}\ \bibnamefont
  {Chen}}, \bibinfo {author} {\bibfnamefont {A.}~\bibnamefont {Kapustin}},
  \bibinfo {author} {\bibfnamefont {A.}~\bibnamefont {Turzillo}},\ and\
  \bibinfo {author} {\bibfnamefont {M.}~\bibnamefont {You}},\ }\bibfield
  {title} {\bibinfo {title} {Free and interacting short-range entangled phases
  of fermions: Beyond the tenfold way},\ }\href@noop {} {\bibfield  {journal}
  {\bibinfo  {journal} {Physical Review B}\ }\textbf {\bibinfo {volume}
  {100}},\ \bibinfo {pages} {195128} (\bibinfo {year} {2019})}\BibitemShut
  {NoStop}%
\bibitem [{\citenamefont {Schnyder}\ \emph {et~al.}(2008)\citenamefont
  {Schnyder}, \citenamefont {Ryu}, \citenamefont {Furusaki},\ and\
  \citenamefont {Ludwig}}]{PhysRevB.78.195125}%
  \BibitemOpen
  \bibfield  {author} {\bibinfo {author} {\bibfnamefont {A.~P.}\ \bibnamefont
  {Schnyder}}, \bibinfo {author} {\bibfnamefont {S.}~\bibnamefont {Ryu}},
  \bibinfo {author} {\bibfnamefont {A.}~\bibnamefont {Furusaki}},\ and\
  \bibinfo {author} {\bibfnamefont {A.~W.~W.}\ \bibnamefont {Ludwig}},\
  }\bibfield  {title} {\bibinfo {title} {Classification of topological
  insulators and superconductors in three spatial dimensions},\ }\href
  {https://doi.org/10.1103/PhysRevB.78.195125} {\bibfield  {journal} {\bibinfo
  {journal} {Phys. Rev. B}\ }\textbf {\bibinfo {volume} {78}},\ \bibinfo
  {pages} {195125} (\bibinfo {year} {2008})}\BibitemShut {NoStop}%
\bibitem [{\citenamefont {Zheng}\ \emph {et~al.}(2020)\citenamefont {Zheng},
  \citenamefont {Achilleos}, \citenamefont {Chen}, \citenamefont {Richoux},
  \citenamefont {Theocharis}, \citenamefont {Wu}, \citenamefont {Mei},
  \citenamefont {Felix}, \citenamefont {Tournat},\ and\ \citenamefont
  {Pagneux}}]{zheng2020acoustic}%
  \BibitemOpen
  \bibfield  {author} {\bibinfo {author} {\bibfnamefont {L.-Y.}\ \bibnamefont
  {Zheng}}, \bibinfo {author} {\bibfnamefont {V.}~\bibnamefont {Achilleos}},
  \bibinfo {author} {\bibfnamefont {Z.-G.}\ \bibnamefont {Chen}}, \bibinfo
  {author} {\bibfnamefont {O.}~\bibnamefont {Richoux}}, \bibinfo {author}
  {\bibfnamefont {G.}~\bibnamefont {Theocharis}}, \bibinfo {author}
  {\bibfnamefont {Y.}~\bibnamefont {Wu}}, \bibinfo {author} {\bibfnamefont
  {J.}~\bibnamefont {Mei}}, \bibinfo {author} {\bibfnamefont {S.}~\bibnamefont
  {Felix}}, \bibinfo {author} {\bibfnamefont {V.}~\bibnamefont {Tournat}},\
  and\ \bibinfo {author} {\bibfnamefont {V.}~\bibnamefont {Pagneux}},\
  }\bibfield  {title} {\bibinfo {title} {Acoustic graphene network loaded with
  helmholtz resonators: a first-principle modeling, dirac cones, edge and
  interface waves},\ }\href@noop {} {\bibfield  {journal} {\bibinfo  {journal}
  {New Journal of Physics}\ }\textbf {\bibinfo {volume} {22}},\ \bibinfo
  {pages} {013029} (\bibinfo {year} {2020})}\BibitemShut {NoStop}%
\bibitem [{\citenamefont {Coutant}\ \emph {et~al.}(2021)\citenamefont
  {Coutant}, \citenamefont {Sivadon}, \citenamefont {Zheng}, \citenamefont
  {Achilleos}, \citenamefont {Richoux}, \citenamefont {Theocharis},\ and\
  \citenamefont {Pagneux}}]{coutant2021acoustic}%
  \BibitemOpen
  \bibfield  {author} {\bibinfo {author} {\bibfnamefont {A.}~\bibnamefont
  {Coutant}}, \bibinfo {author} {\bibfnamefont {A.}~\bibnamefont {Sivadon}},
  \bibinfo {author} {\bibfnamefont {L.}~\bibnamefont {Zheng}}, \bibinfo
  {author} {\bibfnamefont {V.}~\bibnamefont {Achilleos}}, \bibinfo {author}
  {\bibfnamefont {O.}~\bibnamefont {Richoux}}, \bibinfo {author} {\bibfnamefont
  {G.}~\bibnamefont {Theocharis}},\ and\ \bibinfo {author} {\bibfnamefont
  {V.}~\bibnamefont {Pagneux}},\ }\bibfield  {title} {\bibinfo {title}
  {Acoustic su-schrieffer-heeger lattice: Direct mapping of acoustic waveguides
  to the su-schrieffer-heeger model},\ }\href@noop {} {\bibfield  {journal}
  {\bibinfo  {journal} {Physical Review B}\ }\textbf {\bibinfo {volume}
  {103}},\ \bibinfo {pages} {224309} (\bibinfo {year} {2021})}\BibitemShut
  {NoStop}%
\end{thebibliography}%

\end{document}